# Heterostructure and Interfacial Engineering for Low-Resistance Contacts to Ultra-Wide Bandgap AlGaN

Yinxuan Zhu[1,a], Andrew A. Allerman[3], Chandan Joishi[1], Jonathan Pratt[1], Agnes Maneesha Dominic Merwin Xavier[1], Gabriel Calderon Ortiz[2], Brianna A. Klein[3], Andrew Armstrong[3], Jinwoo Hwang[2], Siddharth Rajan[1,2,a]

[1] *Department of Electrical and Computer Engineering, The Ohio State University, Columbus, Ohio 43210, USA*

[2] *Department of Materials Science and Engineering, The Ohio State University, Columbus, Ohio 43210, USA*

[3] *Sandia National Laboratories, Albuquerque, New Mexico 87123, USA.*

**Abstract:**

**We report on the heterostructure and interfacial engineering of metalorganic chemical vapor deposition (MOCVD) grown reverse-graded contacts to ultra-wide bandgap AlGaN. A record low contact resistivity of $1.4 \times 10^{-6}$ $\Omega.cm^2$ was reported on an $Al_{0.82}Ga_{0.18}N$ metal semiconductor field effect transistor (MESFET) by compositionally grading the contact layer from $Al_{0.85}Ga_{0.15}N \rightarrow Al_{0.14}Ga_{0.86}N$ with degenerate doping and proper interfacial engineering considering bandgap-narrowing-induced band offset between channel and contact layer. This represents orders-of-magnitude of lower contact resistivity than that obtained in similar MOCVD-grown structures. A detailed, layer-by-layer analysis of the reverse graded contact and TCAD simulation of the bandgap narrowing effect highlighted that the reverse graded contact layer itself is extremely conductive and interfacial resistance due to bandgap-narrowing-induced barrier between contact and channel dominates the contact resistance.**

[a] Authors to whom correspondence should be addressed

Electronic mail: *zhu.2931@osu.edu, rajan.21@osu.edu*

Ultra-wide bandgap semiconductors have the potential to realize improved electronics for wireless communication and sensing in the mm-wave and terahertz frequency regimes by enabling power amplifiers with high efficiency and gain. Ultra-wide bandgap semiconductors that combine high breakdown electric fields and high saturation velocity are especially suitable for such applications. The high breakdown electric field enables high voltages to be dropped over short distances, thus enabling short drain transit time and high cutoff frequency. The theoretical material limit of power density and gain for a transistor at a given frequency depends directly on the breakdown field and saturation velocity. Over the last two decades, gallium nitride-based high electron mobility transistors (HEMTs) have been demonstrated with excellent large-signal performance, efficiency, gain, and power density approaching theoretical limits [9][14][25]. At lower frequencies, such transistors are thermally limited. However, as frequencies approach the mm-wave and THz (> 100 GHz) regimes, GaN transistors show lower gain, since they are electronically limited by the breakdown field. One potential material candidate for improved performance in the mm-wave and THz regime is ultra-wide bandgap AlGaN. The estimated breakdown field ($F_{BR}$) > 10 MV/cm of Al-rich AlGaN (Al-composition > 50%) is significantly higher than GaN while the estimated saturation velocity ($v_{sat}$>1.5x10$^7$ cm/s) [7] is **similar to** GaN. Therefore, it is expected that AlGaN transistors could enable significantly higher power gain and power density than GaN [21].

Previous work on scaled AlGaN devices has shown promising results but remains limited by contact resistance. High breakdown field (up to 8.5 MV/cm) has been realized in lateral AlGaN/GaN structures [6], and excellent sheet charge and mobility approaching theoretical limits have been obtained in UWBG AlGaN [1][2][3]. While recent work on ultrawide bandgap Al-rich AlGaN channel transistors has enabled near-theoretical electron mobility [1][22], polarization-graded AlGaN FETs [12][20], high current density micro-channel FETs (> 0.9 A/mm), promising large-signal power (2.7 W/mm at 10 GHz) [2], extreme breakdown fields (>8 MV/cm) [8], and high cutoff frequency (> 40 GHz) [3], the high contact resistance has limited device performance in almost all of these designs. In particular, the parasitic RC delays introduced by the high contact and sheet resistance have prevented high frequency current and power gain from reaching the intrinsic material limits of AlGaN. Further optimization of the contact resistance is, therefore, paramount for enhancing the high-frequency performance of this material system.

The high contact resistance to AlGaN is a long-standing challenge and is caused by the low electron affinity, which leads to high metal/semiconductor barrier heights, and the challenge of achieving high doping density. While V-/Zr-based metal stacks have been utilized to achieve ohmic contacts on Al-rich AlGaN channels, the contact resistivities reported are higher than 10$^{-5}$ Ω.cm$^2$ for Al-composition close to 50%, and even higher for Al-content greater than 70% [2][11][16][18]. A promising approach to realize ohmic contacts in this material system is the use of compositionally graded layers (high Al-content AlGaN → low Al-content AlGaN or GaN) along with degenerate n-type doping to compensate for the negative polarization charge resulting from the reverse grade [3][4][10][12]. Low contact resistance of 0.3 Ω.mm ($3\times10^{-6}$ Ω-cm$^2$) to molecular beam epitaxy (MBE) grown 75% AlGaN channels has been previously reported via reverse grading [4]. However, there exist no reports of contact resistance lower than 2 Ω.mm in vertically scaled 2-dimensional and 3-dimensional electron gas channels grown using metal organic chemical vapor deposition (MOCVD).

In this letter, we report on the heterostructure and interfacial engineering of MOCVD-grown reverse graded AlGaN contact layers to achieve record low contact resistances of < 0.3 Ω.mm ($1.4\times10^{-6}$ Ω-cm$^2$) on an $Al_{0.82}Ga_{0.18}N$ n-type channel. We show that highly doped, reverse-graded contact layers graded down to $Al_{0.14}Ga_{0.86}N$ can be grown efficiently using MOCVD. Through an analysis of multiple layers within the reverse-graded stack, we established excellent lateral resistivity for the reverse-graded regions but found that the channel-contact interface creates additional resistance (~ $10^{-6}$ Ω-cm$^2$). We found that introducing a small compositional discontinuity between the channel and reverse-graded contact layers can reduce this interfacial resistance by an order of magnitude, leading to contact resistance of $1.4\times10^{-6}$ Ω-cm$^2$/0.25 Ω-mm, which is by far the best contact resistance achieved to date on such ultra-wide bandgap n-type AlGaN devices.

Three experimental sets are described in this paper. In the first set (A1-A3), three samples with different grade/doping design were investigated. In the second set (B1-B4), different regions of the reverse graded contact were grown to de-embed the contribution of each region. In the last set (C1-C3), we introduced a compositional discontinuity between the reverse graded contact and the channel.

The schematic and energy band diagram for the control n-type AlGaN structure (A1) with a reverse graded layer (calculated using a 1-D Schrodinger-Poisson solver BandEng [26]), are shown in Figure 1(b) and (c). The epitaxial structure consists of 540 nm $Al_{0.85}Ga_{0.15}N$ channel with a silicon doping of $3\times10^{18}$ cm$^{-3}$, a 100 nm highly doped reverse graded contact layer compositionally graded from $Al_{0.85}Ga_{0.15}N$ to $Al_{0.14}Ga_{0.86}N$, and 30 nm highly doped $Al_{0.14}Ga_{0.86}N$ layer. Heavy doping in the reverse graded layer not only compensates for the negative polarization charge but also enables good n$^+$ ohmic metal contact on the $Al_{0.14}Ga_{0.86}N$ surface. Information on the carrier density and mobility are provided in the attached Supplementary Information. The $Al_{0.14}Ga_{0.86}N$ surface (Si ~ $1\times10^{20}$ cm$^{-3}$) displayed a rough surface morphology characterized using atomic force microscopy (AFM) with a root mean square (rms) roughness of 7.5 nm (Figure 2(a)). The layer thicknesses and Al composition were confirmed using high resolution

X-ray diffraction (HR-XRD) (Bruker XRD) measurements, and partially relaxed growth was confirmed using XRD-reciprocal space mapping (RSM) measurements shown in Figure 2(b). The cross-sectional scanning transmission electron microscopy (STEM) image of this sample with energy dispersive X-ray mapping (EDX), as shown in Figure 3(a)-(d), indicates linear grading of Al and Ga in the reverse graded contact layer, which matches our design very well.

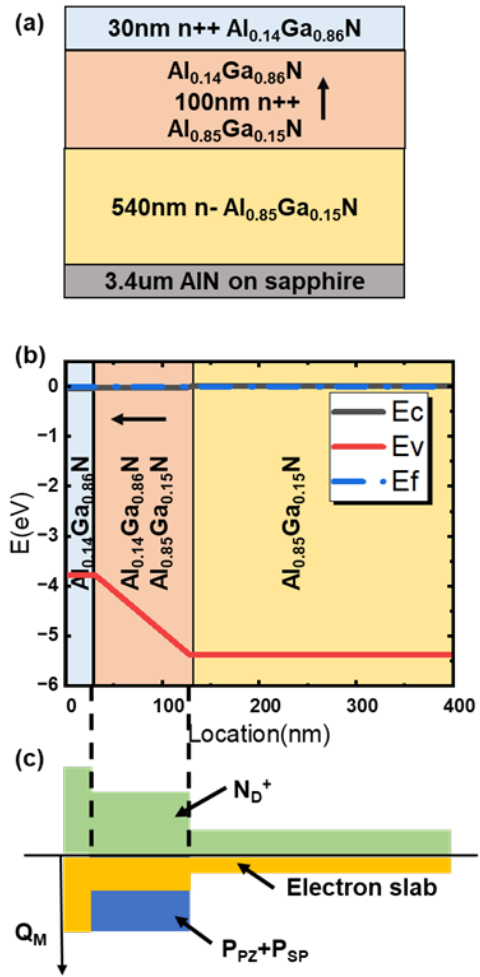

**Figure 1** (a) Epitaxial structure of sample A1, (b) Band diagram of MOCVD-grown MESFET with reverse graded contact layer, and (c) charge profile.

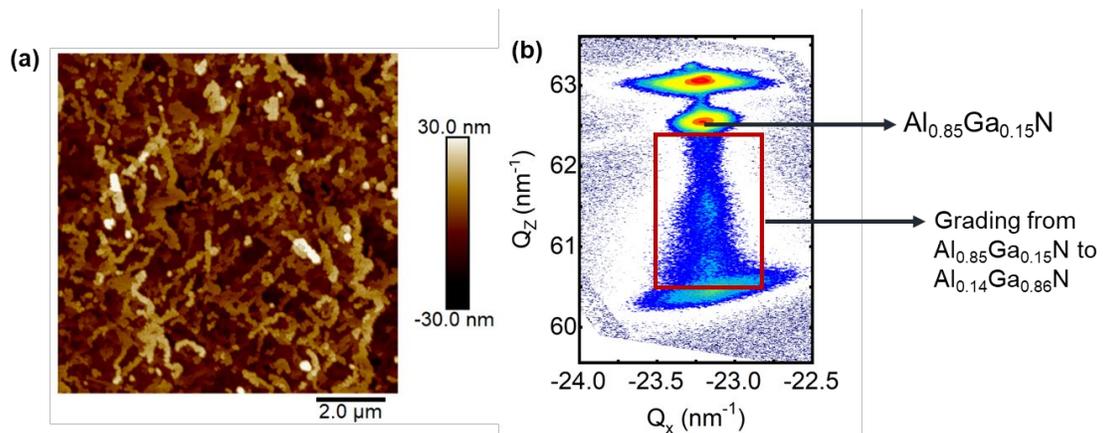

**Figure 2** (a) AFM scan (10um×10um) of the surface of the sample showing an RMS surface roughness of 7.5 nm; (b) XRD reciprocal space mapping data of the epitaxial layer stack.

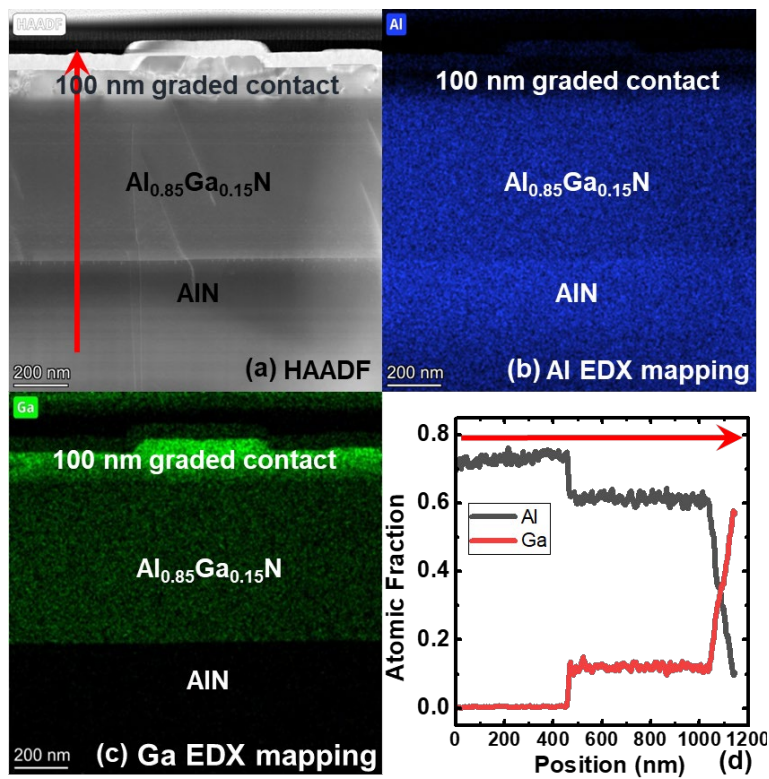

**Figure 3** (a) STEM HAADF image, (b) EDX mapping of Al, (c) EDX mapping of Ga, and (d) Atomic fraction plots of Ga and Al along the direction indicated by red arrow in (a).

Device fabrication commenced with the formation of ohmic contacts using Ti/Al/Ni/Au (20 nm/120 nm/30 nm/100 nm) as the ohmic metal stack deposited by e-beam evaporation. This was followed by rapid thermal annealing at 860 ºC in $N_2$ ambient for 30 s. Thereafter, the devices were mesa isolated and the contact layers in the gate region were recessed using a low-damage $Cl_2$-based dry etch in an inductively coupled plasma reactive ion etching (ICP-RIE) chamber. Similar growth and fabrication processes were also carried out for other 85% AlGaN device in the following experimental sets.

The contact resistances and sheet resistances of sample A1 were extracted via four terminal transfer length measurements (TLM) carried out on a Keysight B1500 semiconductor parameter analyzer. We fabricated and characterized the contact resistances both before (Figure 4(a)) and after (Figure 4(b)) recessed etch of the layers in the gate region. From TLM on the non-recessed contact layers, the metal-semiconductor contact resistance, sheet resistance of the reverse graded contact layer, and the specific metal-to-semiconductor contact resistivity were calculated as 0.023 Ω.mm, 181 Ω/□ and $2.8 \times 10^{-8}$ Ω.cm².

To estimate the specific contact resistivity between the reverse graded layer and the channel layer, we should take into account the finite conductivity of the reverse graded contact layer. A modified equation was used to derive this specific contact resistance to the channel layer [28]:

$$R_{C2} = \frac{\frac{R_{SH\_ch}^2}{R_{SH\_RG}\beta W}\sinh(\beta L_2) + \frac{R'_C \rho_D}{R_{SH\_RG}+\rho_D}(2+\frac{R_{SH\_RG}^2+R_{SH\_ch}^2}{R_{SH\_RG}R_{SH\_ch}}\cosh(\beta L_2))}{\frac{R_{SH\_RG}+R_{SH\_ch}}{R_{SH\_RG}}\cosh(\beta L_2) + \frac{R'_C}{R_{SH\_RG}}\beta W \sinh(\beta L_2)} \quad (1),$$

where $R_{C2}$ is contact resistance, $R_{SH\_RG}$ is sheet resistance of reverse graded contact layer, $R_{SH\_ch}$ is sheet resistance of channel layer, $L_2$ is the distance between metal and recess edge, W is the width of metal contacts, β can be expressed as

$$\beta = \sqrt{\frac{R_{SH\_RG}+R_{SH\_ch}}{\rho_{C2}}} \quad (2),$$

where $\rho_{C2}$ is specific contact resistivity, and $R'_C$ can be expressed as

$$R'_C = \frac{R_{SH\_ch}}{\alpha W} \frac{\cosh(\alpha L_3) + \frac{R_{SH\_ch}}{R_{SH\_RG}+R_{SH\_ch}}\frac{\beta}{\alpha}\tanh(\beta L_4)\sinh(\alpha L_3)}{\sinh(\alpha L_3) + \frac{R_{SH\_ch}}{R_{SH\_RG}+R_{SH\_ch}}\frac{\beta}{\alpha}\tanh(\beta L_4)\cosh(\alpha L_3)} \quad (3),$$

where $L_3$ is the length of metal pad, $L_4$ is the distance between metal and recess edge, and α can be expressed as

$$\alpha = \sqrt{\frac{\rho_{C2}}{R_{SH\_ch}}} \quad (4).$$

The total contact resistance on the recessed structure where the spacings which are used to extract contact resistance and sheet resistance are the lengths of recesses, the sheet resistance of doped $Al_{0.85}Ga_{0.15}N$ channel, and specific contact resistivity estimated by the above equations increased to 0.63 Ω.mm, 3.7 kΩ/□ and $4.38 \times 10^{-6}$ Ω.cm² respectively.

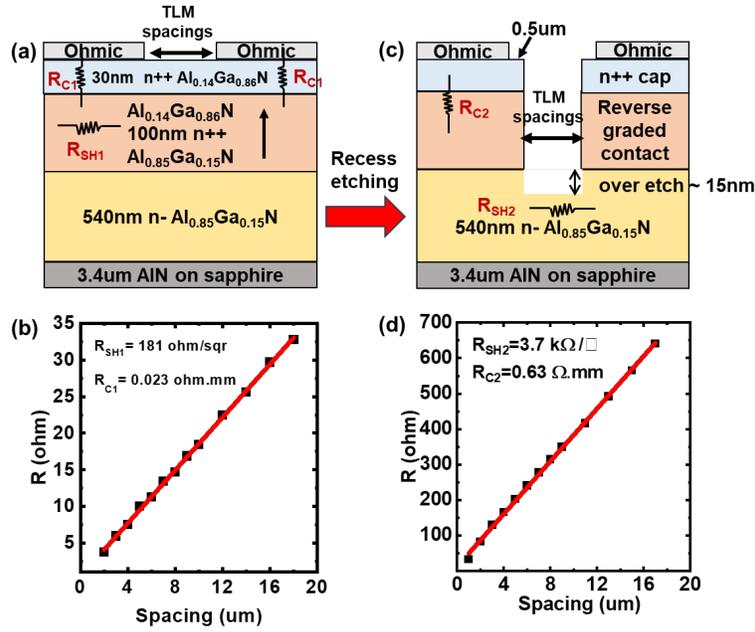

**Figure 4** Sample A1: (a) TLM structures before recessing, (b) TLM fitting plot, extracted M-S resistance and contact layer sheet resistivity, (c) TLM structures after recessing with over etching 15 nm into the $Al_{0.85}Ga_{0.15}N$ channel and (d) Recessed-TLM fitting plot, extracted contact resistance and channel sheet resistance.

To investigate the impact of doping in reverse graded contact and thickness of contact layer, sample A2 and sample A3 were grown. The epitaxial structures and band diagrams are shown in figure 5(b)(c). Sample A2 has the same structure as the one of sample A1, but with lower doping in part of the grading ($Al_{0.85}Ga_{0.15}N$ to $Al_{0.72}Ga_{0.28}N$). The contact resistance ($R_{C2}$) of 1.24 Ω.mm and specific contact resistivity ($\rho_{C2}$) of $3.2\times10^{-5}$ Ω.cm$^2$ were extracted from TLM. Compared with sample A1, Sample A2 has higher contact resistance due to the barrier between contact layer and channel, which stresses the importance of heavy doping in the contact layer.

Sample A3 has thicker reverse graded contact layer (140 nm) compared with sample A1 but with same grading composition as shown in figure 5(c). With thicker reverse graded contact layer and same grading composition, polarization-induced density of holes would decrease, therefore we would ensure all polarization-induced holes are fully compensated with Si dopants. However, $R_{C2}$ of 1.04 Ω.mm and $\rho_{C2}$ of $1.7\times10^{-5}$ Ω.cm$^2$ were extracted from TLM. Thicker contact layers were observed to have rougher morphology as observed in AFM. It is possible that degradation in the microstructure led to an increase in the resistance between 100 nm and 140 nm samples. Therefore, reverse graded contact layer design in sample A1 was used in subsequent experiments.

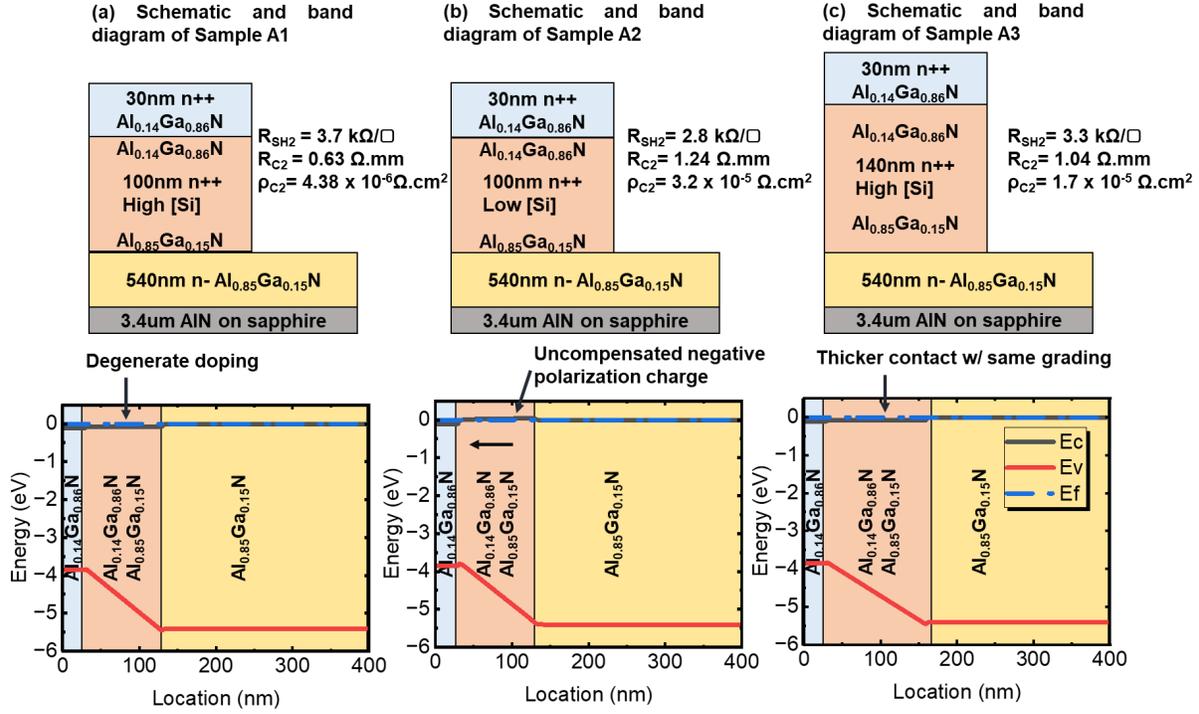

**Figure 5** Schematics and band diagrams of (a) sample A1(with higher Si dopants compared with sample A2), (b) sample A2 (with lower Si dopants in reverse graded contact), (c) sample A3 (with thicker contact layer and same doping level with sample A2).

To understand how each region within the reverse graded contact layer contributes to the total resistance, another sample series was designed. In this series (B1-B4) reverse graded contact layers were grown on the same channel layer and had the same starting composition but different ending compositions (Figure 6(a)). Sheet resistances of 225 Ω/□, 241 Ω/□, 345 Ω/□ and 714 Ω/□ were separately extracted from non-recessed TLM for ending compositions of 15% (Sample B1), 21% (Sample B2), 42% (Sample B3) and 60% (Sample B4), respectively. Here, to yield good contacts on Sample B3 and B4, V/Al/Ti/Au contacts were evaporated by e-beam evaporator instead of Ti/Al/Ni/Au. Since the layers are in parallel, we can de-embed sheet resistance of each layer (RG1 - 15%-21%, RG2 - 21%-42%, RG3 - 42%-60%, RG4 - 60%-85%) from two neighboring layers in parallel. Detailed information on the de-embedding is described in the Supplementary Information. After calculating the sheet resistances for the individual layers, *if we assume isotropic conductivity*, we can estimate the specific vertical resistance associated with each layer according to the equation:

$$\rho_{ISO} = \frac{t}{q\mu_n N_D} = \frac{t^2}{q\mu_n n_s} = R_{SH} * t^2 \qquad (5)$$

where $\rho_{ISO}$, $t$, and $R_{SH}$ are specific vertical resistance, thickness, and sheet resistance of each layer.

From equation (5), we can estimate $\rho_{ISO}$ of $3.14\times10^{-9}$ Ω.cm², $7.03\times10^{-9}$ Ω.cm², $3.94\times10^{-9}$ Ω.cm² and $1.17\times10^{-8}$ Ω.cm² for the layers RG1, RG2, RG3, and RG4 layers, respectively. With this assumption of isotropic transport, the total specific contact resistivity obtained by adding the components sums up to $2.6\times10^{-8}$ Ω.cm², which is much lower than extracted specific contact resistivity from TLM. This discrepancy between the estimated and calculated values suggests that the vertical resistance is not simply the sum of each of the layers, since the measured resistance along the +c direction is significantly higher (~$10^{-6}$ Ω.cm²). We hypothesize that the extra resistance is due to the bottom-most interface (between the N+ contact layer and channel) since that interface is common to all the samples in the series. Furthermore, this interface is the only one that has an abrupt change in doping, which as explained later in this letter, can lead to additional resistance due to bandgap narrowing effects in the heavily doped layer.

To investigate the impact of bandgap narrowing on contact resistance, structures with and without bandgap narrowing were simulated using a 2-dimensional simulation software [27] (figure 8). Bandgap narrowing due to band-tailing effects were calculated using a universal bandgap narrowing model [25] applied to contact layer. The doping in the contact region was assumed to linearly increase from $7\times10^{19}$ cm$^{-3}$ to $1\times10^{20}$ cm$^{-3}$, based on the design and growth conditions used for the contact region. The doping density itself was estimated through separate calibrations carried out for heavily doped regions (Supplementary Information). As shown in the equilibrium energy band diagram, the bandgap narrowing leads to a conduction band offset, and therefore a barrier to electron transport across the channel-contact interface.

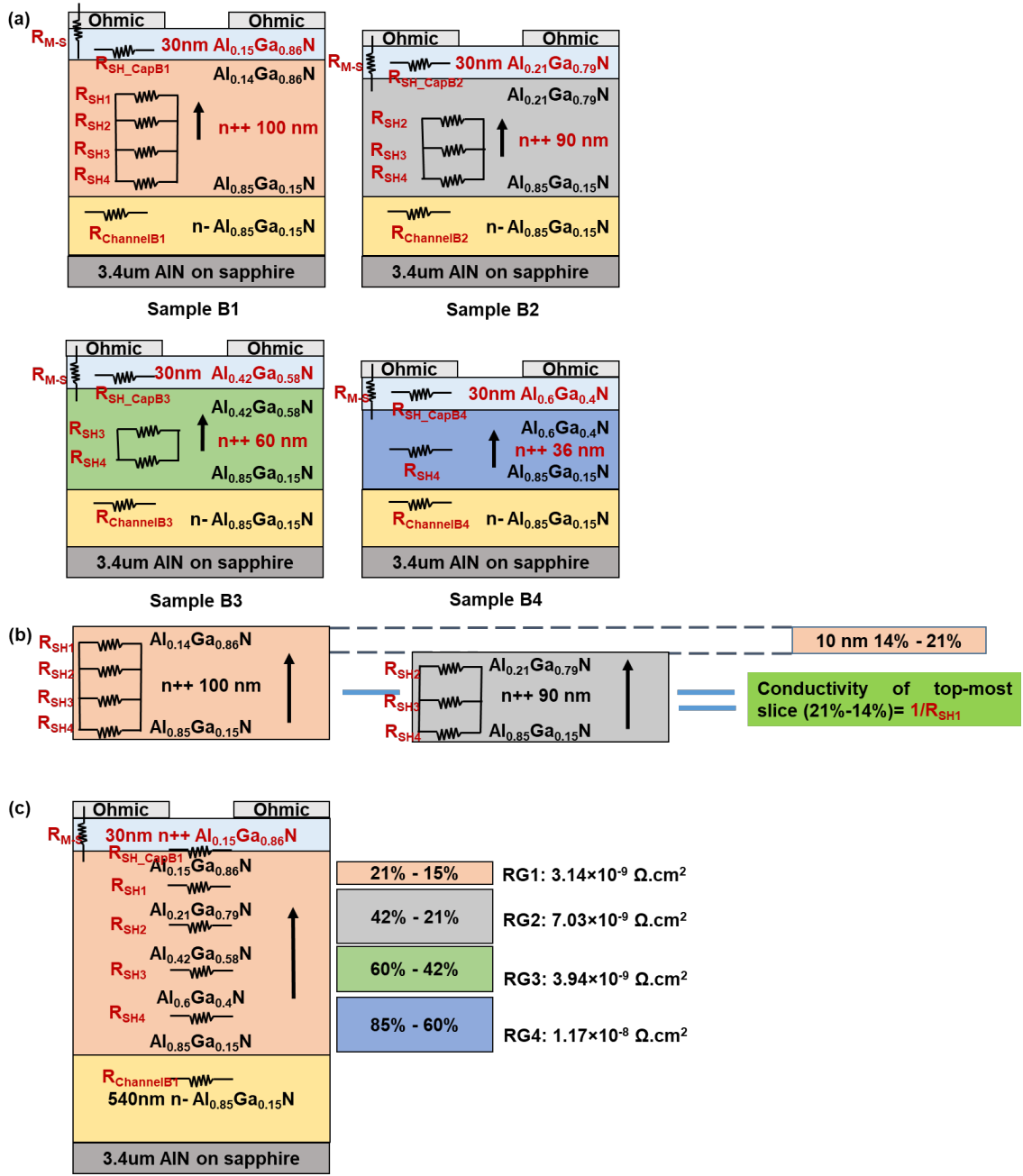

**Figure 6** (a) Samples grown under same grading rate but with different ending Al-composition of 14%, 21%, 42% and 60%, (b) Derivation of sheet resistivity of the first layer – difference of conductivity between sample A and sample B and (c) Specific contact resistivities of each layer.

Two samples with lower channel Al-content (83.5% and 82%) but the same starting Al-content for the contact layer (85%) were designed (figure 7(a)(b)(c)). The idea here was to match the conduction band minima in the 85% heavily doped region (with bandgap narrowing) with a narrower gap channel region. The contact resistance was found to decrease with a reduction in the Al-content in the channel, with contact resistance of 0.25 Ω.mm for 82%, 0.4 Ω.mm for the 83.5% channel, and 0.7 Ω.mm for the compositionally matched 85% channel layer. This shows that lowering the bandgap of the channel layer does in fact lead to a large reduction in contact resistance. In the case of the lowest resistance (82% channel), we estimate a specific contact resistivity of $1.4\times10^{-6}$ Ω.cm² extracted from TLM, implying a 10X reduction in contact resistivity when compared with the uniform bandgap layer.

To understand transport in the presence of this bandgap-narrowing-induced band offset, current-voltage characteristics were also simulated in Silvaco[27]. In addition to drift and diffusion transport, intra-band quantum tunneling, and thermionic emission transport models were included for the interfacial transport across the contact-channel heterojunction. Figure 8(b) shows the simulated two terminal current-voltage curves with and without bandgap narrowing effects. The difference of total resistance extracted from these two IV curves was extracted to be around 0.5 Ω.mm. The simulation results suggest that bandgap-narrowing-induced barriers could indeed explain the experimental data.

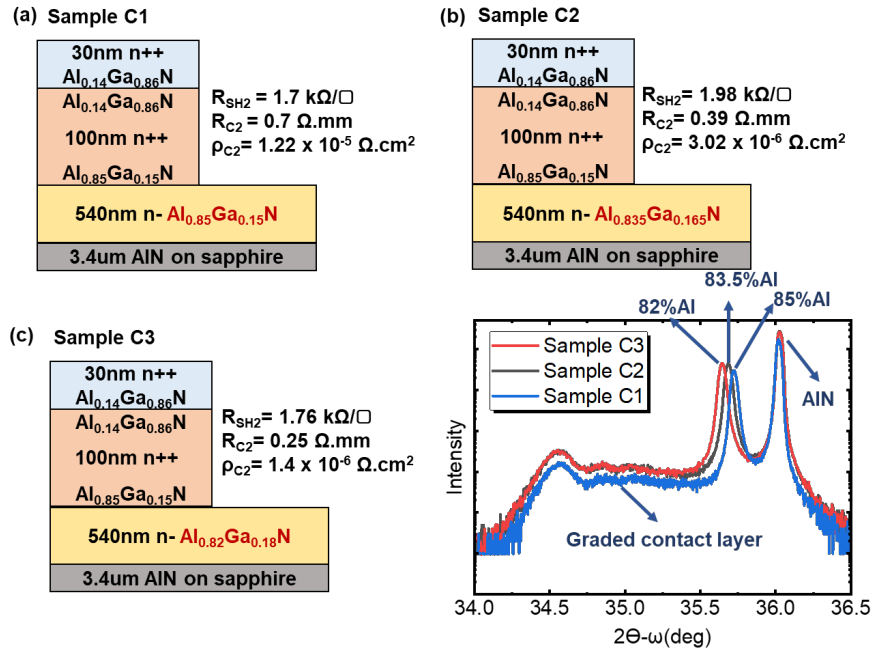

**Figure 7** (a) Sample C1: $Al_{0.85}Ga_{0.15}N$ channel with 100 nm heavily doped contact layer as a control sample; (b) Sample C2: Al-composition of channel is 83.5%, which is slightly lower than Sample C1; (c) Sample C3: Al-composition of channel is 82%, which is slightly lower than Sample C2.

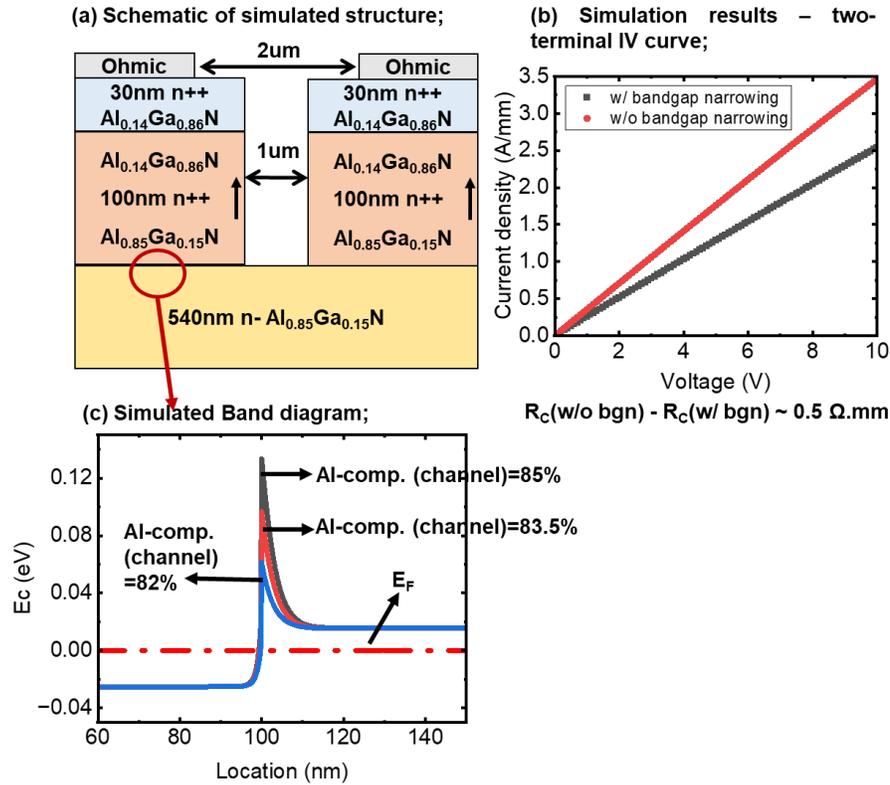

**Figure 8** (a) Schematic of the structures used to simulate the impact of bandgap narrowing; (b) Simulated two-terminal IV curves w/ bandgap narrowing model and w/o bandgap narrowing (bgn) model by 2-D simulation (c) Simulated energy band diagram (conduction band only) by 1-D simulation across the junction between contact layer and channel (considering bgn in the contact layer) indicated by the red circle in (a).

In conclusion, we report on the use of reverse-graded contacts to ultra-wide bandgap AlGaN, enabling contact resistivity as low as $1.4 \times 10^{-6}$ $\Omega.cm^2$ to $Al_{0.85}Ga_{0.15}N$ layers. This represents orders of magnitude lower contact resistivity than that obtained in previously reported MOCVD-grown structures at similar compositions. A layer-by-layer analysis of the total resistance suggested that interfacial layers cause additional resistance, which is hypothesized to be due to bandgap narrowing effects in the heavily doped contact regions. Lower Al-content channel regions were designed to reduce the band offsets introduced by the abrupt change in doping, and this enabled us to achieve contact resistance of $0.25$ $\Omega.mm/1.4\times10^{-6}$ $\Omega.cm^2$. The demonstration here provides a technologically notable approach to realize low-resistance contacts to MOCVD grown Al-rich AlGaN channel transistors, which is a substantial step towards advancing the prospects of this material system in high-frequency electronics.

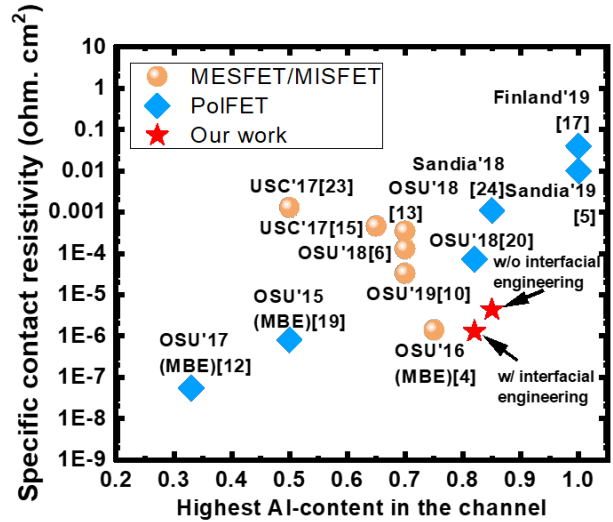

**Figure 9** Reported contact resistivity values in AlGaN PolFETs and MESFETs as a function of the Al-content in the channel.

## Supplementary Material

In this material, we explained the details of de-mebedding sheet resistance of each layer in reverse graded contact layer and MOCVD growth optimization of doping density in the contact layer.

## Acknowledgement

This work was funded by ARO DEVCOM Grant No. W911NF2220163 (UWBG RF Center, program manager Dr. Tom Oder).

# Supplementary Information

**De-embedding sheet resistance of each layer**

Before recessing, metal semiconductor resistances of 0.018 Ω.mm, 0.098 Ω.mm, 0.55 Ω.mm and 1.6 Ω.mm, and sheet resistances of 225 Ω/□, 241 Ω/□, 345 Ω/□ and 714 Ω/□ were separately extracted from TLM results of samples with Al ending composition of 15% (Sample B1), 21% (Sample B2), 42% (Sample B3) and 60% (Sample B4). After recessing, basically with over-etching depth of 15 nm to 20 nm, total contact resistances of 0.63 Ω.mm, 1.05 Ω.mm, 1.3 Ω.mm and 2.25 Ω.mm, and channel sheet resistances of 3.7 kΩ/□, 3.35 kΩ/□, 3.05 kΩ/□ and 3.4 kΩ/□ were extracted from recessed-TLM results of samples with Al ending composition of 15%, 21%, 42% and 60%. Based on TLM results of those four samples, we can split the contact layer into four layers – layer 1(Al-content reversely grading from 21% to 15%), layer 2(Al-content reversely grading from 42% to 21%), layer 3(Al-content reversely grading from 60% to 42%), and layer 4 (Al-content reversely grading from 85% to 60%) as shown figure 6(c). In theory, the extracted sheet resistance of sample A is equal to the sum of the shunt sheet resistances of the $n^{++}$ cap layer, layer 1, layer 2, layer 3, layer 4, and the channel layer as shown in Figure 6(c). Therefore, we have the relation between sheet resistance of sample A and sheet resistances of each layer as,

$$\frac{1}{R_{SH\_B1}} = \frac{1}{R_{SH\_capB1}} + \frac{1}{R_{SH1}} + \frac{1}{R_{SH2}} + \frac{1}{R_{SH3}} + \frac{1}{R_{SH4}} + \frac{1}{R_{ChannelB1}} \quad (S1)$$

where $R_{SH\_B1}$, $R_{SH\_capB1}$, $R_{SH1}$, $R_{SH2}$, $R_{SH3}$, $R_{SH4}$, and $R_{ChannelB1}$ are the extracted sheet resistance from the non-recessed TLM on sample B1, sheet resistance of cap layer in sample B1, sheet resistance of layer 1, sheet resistance of layer 2, sheet resistance of layer 3, sheet resistance of layer 4 and extracted sheet resistance of n-$Al_{0.85}Ga_{0.15}N$ channel layer from the recessed TLM structure on sample B1 which are arranged in a parallel configuration as shown in the figure 6(c).

Similarly, we can derive the relations for sample B2, B3, and B4:

$$\frac{1}{R_{SH\_B2}} = \frac{1}{R_{SH\_capB2}} + \frac{1}{R_{SH2}} + \frac{1}{R_{SH3}} + \frac{1}{R_{SH4}} + \frac{1}{R_{ChannelB2}} \quad (S2)$$

$$\frac{1}{R_{SH\_B3}} = \frac{1}{R_{SH\_capB3}} + \frac{1}{R_{SH3}} + \frac{1}{R_{SH4}} + \frac{1}{R_{ChannelB3}} \quad (S3)$$

$$\frac{1}{R_{SH\_B4}} = \frac{1}{R_{SH\_capB4}} + \frac{1}{R_{SH4}} + \frac{1}{R_{ChannelB4}} \quad (S4)$$

where $R_{SH\_B2}$, $R_{SH\_B3}$, and $R_{SH\_B4}$ are the extracted sheet resistances from the non-recessed TLM on samples B2, B3 and B4 respectively. $R_{SH\_capB2}$, $R_{SH\_capB3}$, and $R_{SH\_capB4}$, are the sheet resistances of the cap layer in Sample B, C and D, whereas $R_{ChannelB2}$, $R_{ChannelB3}$, and $R_{ChannelB4}$ the extracted sheet resistances of n-$Al_{0.85}Ga_{0.15}N$ channel layer from the recessed-TLM on Sample B2, B3 and B4.

If we subtract (2) from (1) while ignoring the difference in the sheet resistances contributed by the heavily doped cap layers compared to the total resistance contributed by the reverse graded layer, we have

$$\frac{1}{R_{SH\_B1}} - \frac{1}{R_{SH\_B2}} = \frac{1}{R_{SH1}} + \frac{1}{R_{ChannelB1}} - \frac{1}{R_{ChannelB2}}, \tag{S5}$$

Therefore, we can derive the sheet resistance of layer 1 as shown in Figure 6(b) by:

$$\frac{1}{R_{SH1}} = \frac{1}{R_{SH\_B1}} - \frac{1}{R_{SH\_B2}} + \frac{1}{R_{ChannelB2}} - \frac{1}{R_{ChannelB1}} \tag{S6}$$

Similarly, we have

$$\frac{1}{R_{SH2}} = \frac{1}{R_{SH\_B2}} - \frac{1}{R_{SH\_B3}} + \frac{1}{R_{ChannelB3}} - \frac{1}{R_{ChannelB2}} \tag{S7}$$

$$\frac{1}{R_{SH3}} = \frac{1}{R_{SH\_B3}} - \frac{1}{R_{SH\_B4}} + \frac{1}{R_{ChannelB4}} - \frac{1}{R_{ChannelB3}} \tag{S8}$$

$$\frac{1}{R_{SH4}} = \frac{1}{R_{SH\_B4}} - \frac{1}{R_{ChannelB4}} \tag{S9}$$

where we ignore the sheet resistances of the cap layer in sample B4 when we calculated $R_{SH4}$. $R_{SH1}$, $R_{SH2}$, $R_{SH3}$, and $R_{SH4}$ are accordingly estimated to be 3136 Ω/□, 781 Ω/□, 684 Ω/□, 902 Ω/□ using equations (1)-(9).

## MOCVD growth optimization of doping density in reverse graded contact layer

Epitaxial structures were grown using metal-organic vapor phase epitaxy (MOVPE) using a Tayio Nippon Sanso SR4000-HT system with trimethylgallium (TMGa), trimethylaluminum (TMAl), ammonia and silane (SiH4). Growth conditions for MESFET layers and reverse grade contact structures were optimized separately for low trap density and high free electron concentration respectively. All structures were grown on previously prepared 3.4 um thick AlN epi layers grown on c-plane, sapphire substrates mis-cut 0.25° toward the m-plane.

Room temperature Hall effect measurements using Van der Pauw method were used to determine the free electron concentration in 230-300 nm thick Si-doped, AlGaN calibrations layers at various compositions between Al-composition= 0.85 and 0.14. Contacts with linear current voltage characteristics were easily obtained using Indium preforms pressed onto the epilayer surface due to the high free electron concentrations achieved. Table 1 shows the maximum electron concentration from optimization of the Si level at Al mole fractions used to construct a piecewise linear grade between the MESFET layer (x=0.85) and the contact layer (x=0.14). The ending composition of $Al_{0.14}Ga_{0.86}N$ was selected over GaN as the electron concentration of $1.7 \times 10^{20}$ cm$^{-3}$ is sufficiently high that the metal-semiconductor contact resistance was only 0.023 ohm.mm and is a negligible contribution to the total contact resistance. Thus, extending the grade down to GaN would result in more surface roughening due to relaxation from increased compressive strain without a meaningful reduction in contact resistance. The reverse grade of Al composition was fixed at 100 nm but the thickness of the ending $Al_{0.14}Ga_{0.86}N$ layer was varied in this study.

| Al-comp. | Carrier density (cm$^{-3}$) | Mobility (cm$^2$/V.s) |
|---|---|---|
| 0.85 | 1.6 x 10$^{19}$ | 26.1 |
| 0.72 | 6.7 x 10$^{19}$ | 35.2 |
| 0.6 | 8.5 x 10$^{19}$ | 35.8 |
| 0.42 | 8.0 x 10$^{19}$ | 33.6 |
| 0.21 | 1.11 x 10$^{20}$ | 58.2 |
| 0.14 | 1.7 x 10$^{20}$ | 44.7 |

**Table S1** Room temperature Hall effect measurements of calibration layers at Al mole fractions used to construct a piecewise linear grade between the Al mole fraction of the MESFET (x=0.85) and the contact layer (x=0.14).